\documentclass[manuscript]{emulateapj}

\usepackage{natbib,color,graphicx}
\usepackage{ textcomp }
\usepackage{float}
\usepackage{soul}
 \usepackage{sidecap}

\shorttitle{Solar Eruption 2013 November 06}
\shortauthors{Zucca et al.}

\begin{document}

\title{Understanding CME and associated shocks in the solar corona by merging multi wavelength observations}


\author{P. Zucca $^{1, 2}$, M. Pick $^{2}$, P. D{\'e}moulin$^{2}$, A. Kerdraon$^{2}$, A. Lecacheux $^{2}$,P.T. Gallagher $^{1}$}

\affil{$^{1}$ School of Physics, Trinity College Dublin, Dublin 2, Ireland. }
\affil{$^{2}$ LESIA, Observatoire de Paris, CNRS, UPMC, Universit{\'e} Paris-Diderot, 5 Place Janssen, Meudon 92195, France.}




\begin{abstract}

Using multi-wavelength imaging observations, in EUV, white light and radio, and radio spectral data over a large frequency range, we analyzed the triggering and development of a complex eruptive event. This one includes two components, an eruptive jet and a CME which interact during more than 30 min, and can be considered as physically linked.  This was an unusual event.

 The jet is generated above a typical complex magnetic configuration which has been investigated in many former studies related to the build-up of eruptive jets;  this configuration includes fan-field lines originating from a corona null point above a parasitic polarity, which is   embedded in one polarity region of large Active Region (AR).

The initiation and development of the CME, observed first in EUV, does not show usual signatures.  In this case, the eruptive jet is the main actor of this event. The CME appears first as a simple loop system which becomes destabilized by magnetic reconnection between the outer part of the jet and the ambient medium. The progression of the CME is closely associated with the occurrence of two successive types II bursts from distinct origin.

An important part of this study is the first radio type II burst for which the joint spectral and imaging observations allowed: i) to follow, step by step, the evolution of the spectrum and of the trajectory of the radio burst, in relationship with the CME evolution; ii) to obtain, without introducing an electronic density model, the B-field and the Alfv\'en speed.

\end{abstract}





\section{Introduction}
\label{Sect:Introduction}

Coronal mass ejections (CME) are large-scale energetic events associated with various manifestations of solar activity (e.g., flares, eruptive prominences, shocks). The correlation between the kinematics of the CMEs with these different forms of solar activity has been, for several decades, a major tool to shed light into the physical mechanisms of CME development.

CMEs have been frequently observed in white light coronagraph images as having a so called three-part structure, consisting of a bright rim surrounding a dark void which contains a bright core \citep{Illing1985}.  The SOHO/LASCO and more recently STEREO observations showed that CMEs are consistent with a two-dimensional projection of a three-dimensional magnetic flux rope \citep{Chen1997,Chen2000,Thernisien2006,Thernisien2009}.  The authors concluded that the cavity, seen in white light, can be interpreted as the cross section of an expanded flux rope. \citet{Vourlidas2013} gave arguments implying that at least 40\% of the observed CMEs have flux-rope structures.
  
In recent years, new prominent results  on CME initiation mechanisms and their early development in the low corona have arisen from  EUV observations with the EUV Imager of the STEREO/SECCHI telescope \citep[EUVI;][]{Wuelser2004} and  from the Atmospheric Imaging Assembly on board the \emph{Solar Dynamic Observatory} \citep[$SDO$/AIA;][]{Lemen2012}.
\citet{Patsourakos2010} showed that the CME formation is first characterized by slow, self-similar, expansion of slowly-rising loops, possibly triggered by a rising filament, that leads to the formation of a bubble-shaped structure within about 2 minutes. This is consistent with the transformation, by magnetic reconnection, of loops into a flux rope structure as predicted by several models \citep[e.g.,][]{Lynch2008}.  The AIA multi-temperature observations have given access to detailed description  of a CME namely: i) the ejection of a plasma blob transforming rapidly into a growing hot flux rope that stretches the upper field lines; ii) the appearance of a Y-type magnetic configuration at the bottom of the flux-rope, in which a bright thin line (i.e., a Current Sheet, CS) extends downward; and iii) the shrinkage of magnetic field lines observed underneath the CME \citep{Cheng2011, Cheng2013}.  All these above observations are consistent with the CME eruption model proposed by \citet{Lin2004}. This model is based on a flux rope magnetic configuration overlying a photospheric polarity inversion line. This flux rope becomes unstable and erupts building up behind a CS, which convert the surrounding B-field in a new poloidal around the flux rope. In radio, the formation and development of reconnecting CS behind an erupting flux rope was imaged by the Nan\c cay radio-heliograph \citep{Pick2005,Huang2011,Demoulin2012}.

CMEs are frequently associated with type~II radio bursts which are a signature of a shock formation and propagation in the corona at speeds higher than those of the  Alfv\'en speed. These bursts are generated by shocks exciting Langmuir waves which decay into radio waves at the local plasma frequency and/or its harmonics \citep[see e.g.,][]{Melrose1980}. 
   A long debate on the physical mechanisms which generate these shocks is still ongoing \citep[see, e.g.,][]{Vrsnak2008, Vasanth2011}. \citet{Nindos2011} has led to the conclusion that coronal shocks may be generated by two different mechanisms: blast-waves initiated by the plasma pressure of a flare, and piston driven shocks due to CMEs. Several statistical studies on the association of CMEs with type~II radio bursts can be found in the literature \citep[see, e.g.,][]{Gopalswamy2009}. \citet{Ramesh2012} have found that 92\% of the type~II bursts observed at 109~MHz are associated with CMEs and are located near their leading edge. However, the sources of the coronal type~II were often found to be located not in front but on the flanks of CMEs, \citep[see, e.g.,][]{Classen2002,Cho2007,Demoulin2012,Zucca2014}. 

Coronal type~II bursts were also often observed jointly with the occurrence of EUV waves, which are large-scale, bright, wave-like disturbances visible in EUV. Several authors have recently taken advantage of the high cadence observations and of the simultaneous dual (or sometimes triple) view-points obtained with STEREO/EUVI, $SDO$/AIA and PROBAB2/SWAP \citep{Berghmans2006} instruments to study the association between CMEs and EUV waves \citep[see, e.g.,][]{Wang00,Gallagher2011}. For instance, \citet{Veronig2010} found that the development of an EUV wave exhibits two phases: a first phase consistent with a wave driven by the expanding flanks of the CME \citep[e.g.,][]{Carley2013}, and a second one where the wave propagates freely. However, the physical nature of these waves and their association with type~II bursts is still unclear and no single model can account for the large variety of EUV waves observed \citep{Warmuth2010,Zhukov2011}.

To understand the nature of the shock and its association with CMEs, EUV waves, and flares, detailed studies of the complex morphology present in radio burst spectra are required. This complex morphology shows up, for instance, under the form of a splitting of the emission bands into two lanes \citep{Smerd1974,Vrsnak2001}, or a fragmentation and an abrupt change of their drift rates \citep[see, e.g.,][]{Pohjolainen2008,Kong2012}. These various morphologies are related to the characteristics of the eruption and to the properties of the surrounding corona in which the shock is propagating. In particular, the electron density and the magnetic field characterize the ambient medium, which then determine the Alfv\'en speed. This characteristic speed is important for the formation of a shock and for the conditions under which the radio burst can be initiated. Furthermore, both the coronal density and the magnetic field configuration are crucial to determine the radio burst frequency drift and its duration. 

While numerous studies have been realized on the origin of the shocks and their association with CMEs, there are in fact a very small number of cases for which it has been possible to study such events simultaneously through radio spectra obtained on a large frequency scale, and through radio and EUV images obtained with a high enough time resolution to follow their evolution in detail.
    To contribute to the understanding of when and where the CMEs and coronal shocks are produced and how they relate to type~II bursts and EUV waves, we study here the complex morphological spectral features of a radio event observed on 06 November 2013 together with imaging EUV and radio observations.

The November 06 event includes two components, an eruptive jet and a CME, which interact during more than 30 min, and can be considered as physically linked.  The magnetic configuration, in which eruptive jets are produced, has already been studied with several magneto-hydrodynamic (MHD) numerical simulations \citep[e.g.][]{Pariat2009,Pariat2010}. Conversely, eruptive events, such as the 06 November one which was accompanied by a CME, are not frequently described in the literature.
 
The results of the  data analysis presented here  take advantage  from particularly favorable conditions: i) same field of view on EUV and radio instruments; ii) joint radio spectral (0.5-1000 MHz) and multi-frequency imaging observations (150-450 MHz) at high cadence (better than 1s) and with an high sensitivity; iii) a broad-band frequency spectrogram obtained by the combination of different spectrographs.

 We identify step by step the causes of the type~II spectral fragmentation in relationship with the CME evolution and the ambient medium.  We obtain for each step, without introducing an electronic density model or a MHD simulation, the upstream plasma density, the Alfv\'en Mach number for the shock and the magnetic strength. 
  The end of this type~II burst is followed, several minutes later, by a second type~II burst of shorter duration. In the absence of imaging observations, the spectral versus time evolution would have led us to conclude to a reactivation of the original type~II burst. We will show that this is not the case.

The paper is organized as follows:  Section~\ref{Sect:Joint} provides first a  description of the observations, and then we present the data analysis which mainly includes: i) a brief overview of the radio event properties; ii) the magnetic configuration of the active region and of its environment; iii a detailed joint EUV and  radio  analysis of the CME and of the associated type II bursts. Section~\ref{Sect:Shock} presents the method through which the observations of the first radio type II burst leads to an estimation of physical parameters such as the Alfv\'en velocity, the density and the magnetic field of the ambient medium. In  Section~\ref{Sect:Discussion}, we discuss what we learnt on: i) the r\^{o}le of the eruptive jet and of the ambient medium for setting up the CME; ii) the nature of the two shocks, associated with two radio type II bursts, which occur during the CME progression. The main findings are summarized in Section~\ref{Sect:Conclusions}.

\section{Joint evolution of the EUV, white light, and radio emissions}
\label{Sect:Joint}

\subsection{Observations}
\label{Sect:Observations}

\begin{figure*}
\begin{center}
\IfFileExists{fastCompil.txt}{
\includegraphics[width=0.8\textwidth,clip]{Figure1}	                
	                        }{
\includegraphics[trim= 3cm 2cm 3cm 4cm,clip=true,width=11cm,angle=90]{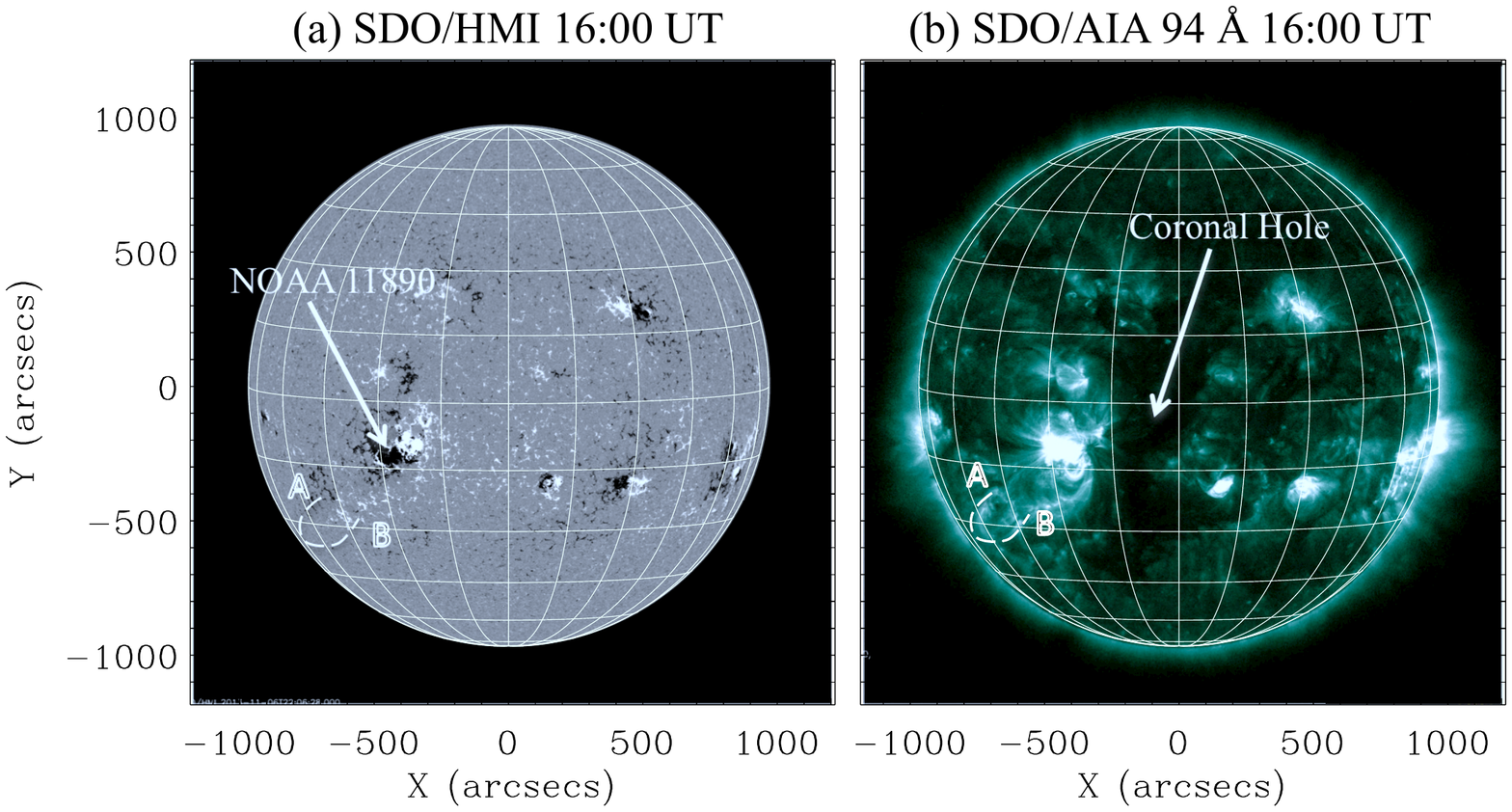}
}
\caption{(a) \textit{Left Panel}: the $SDO$/HMI magnetogram on 2013 November 6; the active region NOAA 11890 is indicated with a white arrow; A and B indicate two regions of opposite polarity (see Section~\ref{Sect:EUV_type_III}). (b) \textit{Right Panel:} $SDO$/AIA 94 \AA\ image at 16:00 UT; the extended equatorial coronal hole, indicated by an  white arrow, is located on the side of the active region.}
\label{Figure1}
\end{center}
\end{figure*}

A GOES M3.8 class flare started on 2013 November 06 at 13:39 UT in the active region (AR) NOAA 11890 (S12 E35).
 The flare maximum occurred at 13:46 UT and an associated coronal mass ejection (CME) was observed at low altitude by $SDO$/AIA starting to rise at $\sim$13:44:00 UT. It was also later observed at $\sim$14:36 UT with the Large Angle and Spectrometric Coronagraph on board the \emph{Solar and Heliospheric Observatory} \citep[$SOHO$/LASCO;][]{Brueckner1995}. An $SDO$/HMI magnetogram and a $SDO$/AIA EUV image at 94~\AA\ are displayed in Figure~\ref{Figure1}; NOAA 11890 was located on the eastern side of an extended coronal hole. 


 Spectral radio observations were obtained with different instruments  (Figure~\ref{Figure2}): the radio spectrograph ORFEES (Observation Radio Frequence pour l'Etude des Eruptions Solaires) which is a new radio-spectrograph located in Nan\c cay and observing between 140 and 1000~MHz, the e-Callisto spectrograph at the Rosse Solar-Terrestrial observatory \citep[RSTO;][]{Zucca2012}, the Decametric Array in Nan\c cay \citep[DAM;][]{Lecacheux2000} observing between 70 and 30~MHz, and the Wind WAVES spectrograph \citep{Bougeret1995} observing between 13.825 and 1.075 MHz. Radio Imaging was obtained with the Nan\c cay radio-heliograph \citep[NRH;][]{Kerdraon1997} which observed at 9 different frequencies between 445 and 150~MHz on November 06, 2013.

\subsection{Overview of the radio event}
\label{Sect:Overview}

During the hours preceding the onset of the event, the main activity, in the eastern hemisphere, consisted in a noise storm \citep{Elgaroy1976} which was observed  in the whole frequency range of the NRH. This noise storm is located south east of AR 11890, and has a negative circular polarization. We will name hereafter negative, respectively positive polarization, the polarization of the  ordinary mode in a negative, entering in the photosphere (respectively positive, going out the photosphere) magnetic field. For the present noise storm, which is supposed to be emitted in the ordinary mode, and is located in large scale negative magnetic fields, the negative polarization is what is expected \citep{Elgaroy1976} . We shall also notice that type II and type III radio bursts are expected to have the polarization of the ordinary mode.

\begin{figure*}
\begin{center}
\IfFileExists{fastCompil.txt}{
\includegraphics[width=0.8\textwidth,clip]{Figure2}	                
	                        }{
\includegraphics[trim= 1cm 4cm 1cm 2cm,clip=true,width=13cm,angle=0]{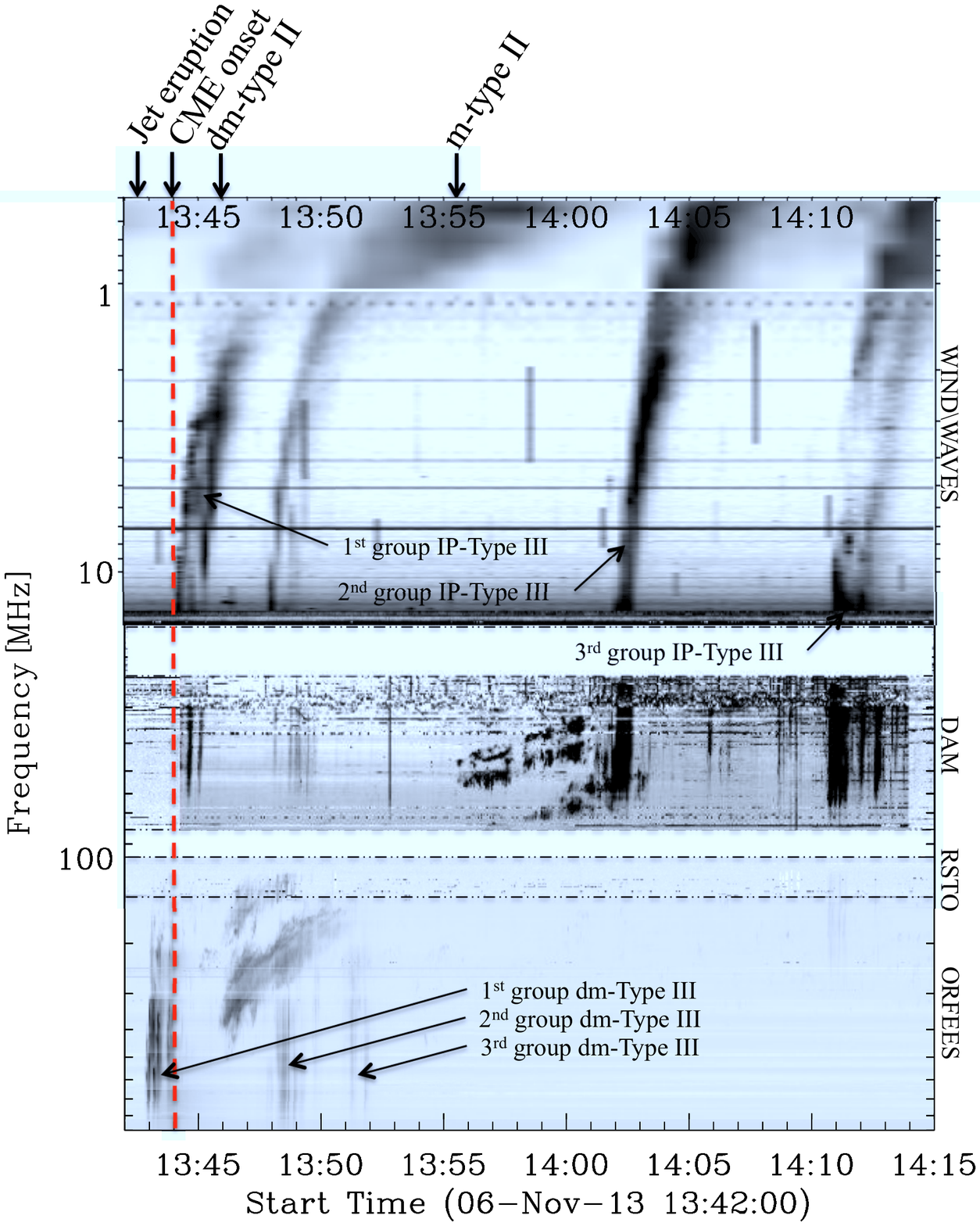}
}
\caption{Dynamic Spectrogram of the event. The frequency range, from 900 to  0.5~MHz, is covered by ORFEES (900-140~MHz), e-Callisto(140-100~MHz), DAM (90-25~MHz), and WIND/WAVES (16-0.5~MHz). The first group of  dm-type~III bursts starts at 13:42:58 UT, while the first interplanetary type~III bursts start at 13:43:57 UT. The red dashed line indicates the separation in time between these two groups. A  metric type~II radio burst starts at 13:45:59 UT then faints; a decameter type~II burst is observed after 13:55 UT  and is  followed by two IP type~III burst groups.} 
\label{Figure2}
\end{center}
\end{figure*}

  An overview of the development of the radio event is shown in Figure~\ref{Figure2}, which displays a  synthetic spectrum of the event obtained by combining the data from the different spectrographs.

 This first radio emission is a group of decimetric (dm) type~III bursts starting at 13:42:58 UT. These bursts are observed only at frequencies higher than 100~MHz and they end at 13:43:57 UT. They are followed  by  interplanetary (IP) type~III bursts, starting around 70~MHz, approximately at the time when the CME observed by $SDO$ starts to rise. The red dashed line in Figure~\ref{Figure2} marks the transition time between the dm and IP type~III bursts. Two other groups of dm type~III bursts are recorded later. 

The first group of IP type~III bursts is followed at 13:45:59 UT by the onset of a type~II radio burst. This burst shows both the  fundamental (F) and harmonic (H) emission and also a band splitting particularly visible in the harmonic emission. The F and H emissions fade in the spectrum at respectively 13:49:00 UT and 13:51:00 UT and are observed to start again at $\sim$13:55:30 UT, respectively at $\sim$45~MHz (F) and $\sim$80~MHz (H). They end at $\sim$14:02~UT, in coincidence with the occurrence of a second group of IP type~III bursts (see  Figure~\ref{Figure2}). A last group of IP type~III bursts is observed at $\sim$14:10:30 UT. 


\begin{table}[h]
\centering
\begin{tabular}{clcc}
\hline
Event           & Movie             & time           & number  \\ 
\hline
SDO/AIA 171 \AA & Direct            & 13:30-14:20 UT &  1 \\
SDO/AIA 171 \AA & Run-Diff.         & 13:40-14:20 UT &  2 \\
SDO/AIA 193 \AA & Run-Diff. and NRH & 13:45-13:52 UT  &  3 \\
SDO/AIA 193 \AA & Run-Diff.         & 13:30-14:30 UT &  4   \\
SDO/AIA 131 \AA & Run-Diff.       & 13:30-14:30 UT & 5 \\ 
\hline
\end{tabular}
\caption{List of available movies}
\label{Table:movies}

\end{table}

\subsection{Magnetic configuration of the active region and of its environment}

\begin{figure*}
\begin{center}
\includegraphics[trim= 1cm 6cm 1.5cm 4cm,clip=true,width=14cm,angle=90]{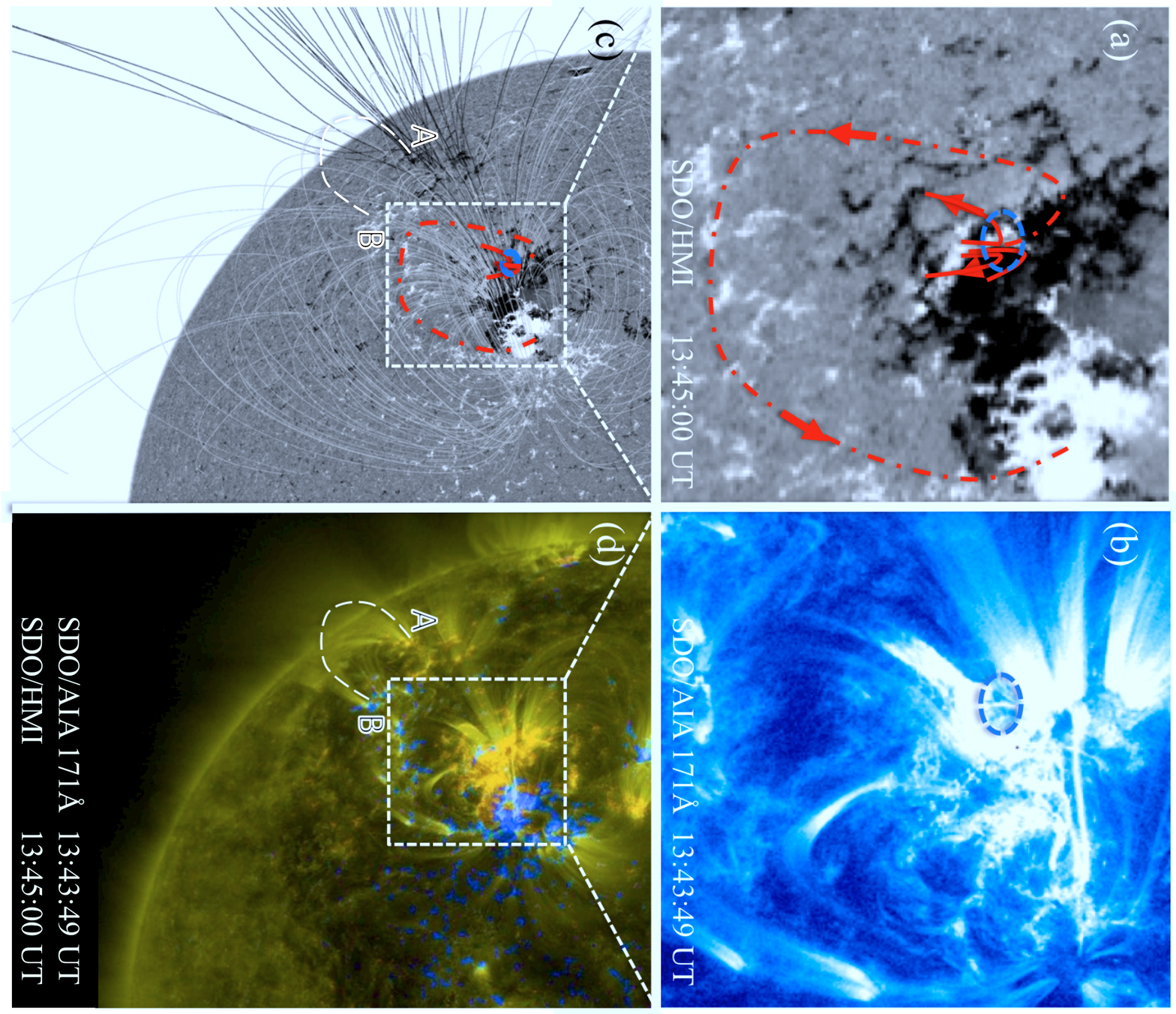}

\caption{Magnetic configuration and EUV images of the source region. (a) $SDO$/HMI magnetogram at 13:45:00 UT and (b) $SDO$/AIA 171 \AA\ for the same field of view at 13:43:49 UT. The circular ribbon flare is indicated with a blue dashed line, while the magnetic field line connecting the null point to the positive polarity is indicated in red. (c) PFSS extrapolation superposed on the $SDO$/HMI magnetogram at 13:45 UT. (d) $SDO$/AIA 171 \AA\ emission superposed on the  $SDO$/HMI magnetogram (positive polarity displayed in blue). The smaller loop structure with foot-points labelled A and B is also shown (see Section 3.2).}

\label{Figure3}
\end{center}
\end{figure*}

To understand the successive phases of the eruption, it is necessary to determine first the magnetic configuration of this AR and of its environment;  we deduce it from the $SDO$/HMI and $SDO$/AIA 171 \AA. In Figure~\ref{Figure3}, the $SDO$/HMI magnetogram is shown (Panel a) together with the $SDO$/AIA EUV image at 171 \AA\ (Panel b). Inside the eastern trailing region of negative polarity, we note the presence of an embedded small positive and parasitic polarity. Using a potential field extrapolation and a three-dimensional MHD numerical simulation, \citet{Masson2009} showed, in a similar case, that the active region includes fan-field lines originating from a coronal null point. This structure has the shape of a dome with the null point at its top.  
\citet{Masson2009} linked the circular shape of the observed  flare ribbons  with the photospheric mapping of the fan-field lines. The ribbon brightening would be due to the chromospheric impact of the particles accelerated near the null point by reconnection between the field lines located just below and above the fan.


By analogy with the Masson et al. event, in our case, the coronal structure is  evidenced in Figure~\ref{Figure3}a and Figure~\ref{Figure3}b by, a blue dashed circle and by the red short field lines. Figure~\ref{Figure3}c displays the  $SDO$/HMI magnetogram together with the potential-field source-surface (PFSS) extrapolation \citep{Schatten1969} using the software of  \citet{Schrijver2003}\footnote{http://www.lmsal.com/$\sim$derosa/pfsspack/}. We also note the presence of more than one embedded (parasitic) polarity observed in the trailing spots; the structure might then be more complex than the one described by \citet{Masson2009}. A coronal loop system is also plotted in Figure~\ref{Figure3}c and Figure~\ref{Figure3}d. It is  anchored in two regions of opposite polarity A and B, south-east of the active region. 

\subsection{EUV eruptive jet and type~III bursts}
\label{Sect:EUV_type_III}

The eruption took place above the active region (AR) NOAA 11890 which was classified as a $\beta\gamma\delta$ region. The eruptive jet is first detected at $\sim$ 13:41:48 UT, in the 171 \AA~ channel of SDO/AIA, as a thin ascending structure. A sudden brightening appears at its basis at 13:42:12 UT, resulting in an increased lateral size. The jet then shows a more complex shape with different branches visible at 13:44:37 UT, 13:47:02 UT and later (see Movie 1). A brightening, visible on running differences (see Movie 2) appears above the main body of the jet at 13:45:36. The jet lasts until 14:10 UT, beginning to  turn and move downward after 14:00 UT, pointing toward a belt of positive magnetic polarities located south of AR 11890.

\subsubsection{Decimeter (dm) type~III burst groups}

\begin{figure*}
\begin{center}
\IfFileExists{fastCompil.txt}{
\includegraphics[width=\textwidth,clip]{Figure4}	                
	                        }{
\includegraphics[trim= 1cm 4cm 1cm 3cm, width=13cm, clip=true,angle=0]{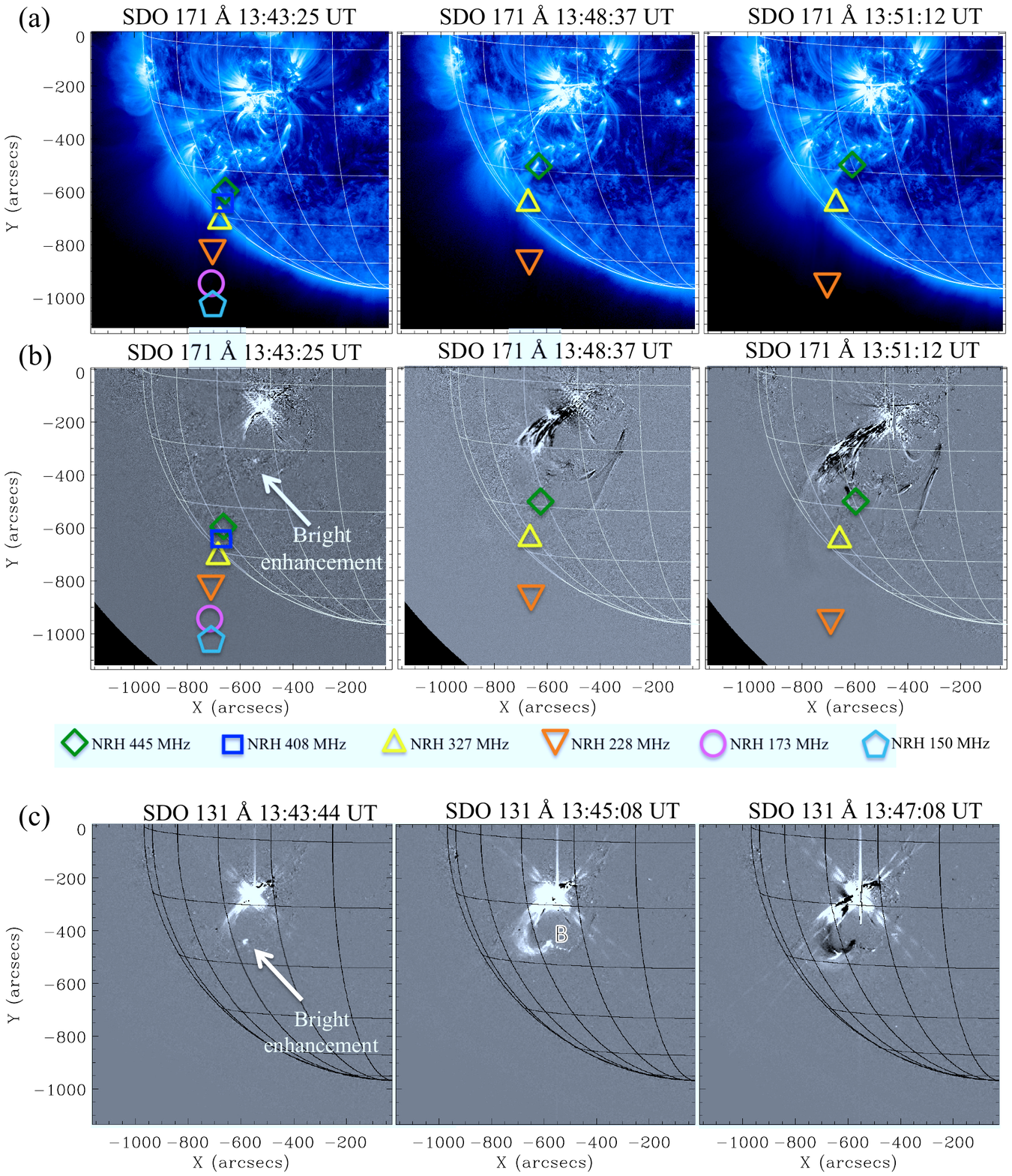}
}
\caption{(a) Erupting phase sequence showing the EUV emission observed with $SDO$/AIA 171 \AA\ at the time of the three successive groups of dm-type~III bursts indicated in Figure~\ref{Figure2}; the location of the dm-type III bursts observed at 13:43:01 UT, 13:48:13 UT and 13:51:25 UT with the NRH are indicated with the colored symbols from 445 MHz to 150 MHz.(b) Running difference sequence of $SDO$/AIA images at 171 \AA\~ ; the white arrow shows the EUV bright source which is located at the bottom of the trajectory traced by the electron beams producing the type III burst emission.(c) Running difference sequences of $SDO$/AIA images at  131 \AA\ showing the bright bridge established between the jet and the  bright source; note that this source is located along field lines anchored in the  B region of positive polarity.}
\label{Figure4}
\end{center}
\end{figure*}

 The  sources of the three dm type~III burst groups were imaged every second at different frequencies by the NRH. All of them have \textit{positive  polarization}. Such polarization means that they are emitted along magnetic field lines emerging from the photosphere. Their locations, measured respectively at 13:43:06 UT, 13:48:37 UT and 13:51:12 UT, are reported, for each group, in the three  direct and running difference images of SDO/AIA at 171 \AA~ displayed in  Figure~\ref{Figure4}a. For the two last groups, because of the progression of a type~II burst, simultaneously detected by the NRH (Figure~\ref{Figure2}), the source locations could be determined only at frequencies higher  than 150 MHz.  These locations, which are nearly identical for the three groups, trace the path followed by the electron beams responsible for the type~III radio emission (see also Figure~\ref{Figure3}c). The three groups have starting frequencies higher than 1 GHz, which implies an acceleration region located low in the corona.
 

The first group occurs soon after the sudden broadening near the base of the jet (see Figure~\ref{Figure4}a), and at the time when a EUV bright and compact EUV source  starts to be observed above the region of positive polarity near B (see  the arrow). Some elongated thin brightenings are also observed in the same region (see in particular Movie 1 at 171 \AA~ and  Movie 5 at 131\AA). Soon after, a bright "like-bridge shape" appears between this bright source and the western part of the jet. It is particularly clear in the run-difference  images at 131 \AA. This observation suggests that this bridge and the EUV bright source result from the reconnection  between  the western side of the eruptive jet and field lines anchored in this region of positive polarity. A few weaker bright points also appear near this main EUV source, which  suggests that other field lines undergo  a similar reconnection process. This process will also accelerate the electrons responsible for the radio type III bursts. This is consistent with the trajectory of the electron beams (revealed by the positions of the radio sources) which lies precisely above the EUV bright source. Furthermore the sign of polarization of these type III implies that they originate above a region of positive polarity which is the case. However, we note that no field lines above this region are observed in Figure~\ref{Figure3}c on the PFSS map. 

\subsubsection{The first interplanetary type~III bursts}

\begin{figure*}
\begin{center}
\IfFileExists{fastCompil.txt}{
\includegraphics[width=0.8\textwidth,clip]{Figure5}	                
	                        }{
\includegraphics[trim= 1cm 3cm 1cm 3cm,clip=true,width=14cm,angle=-90]{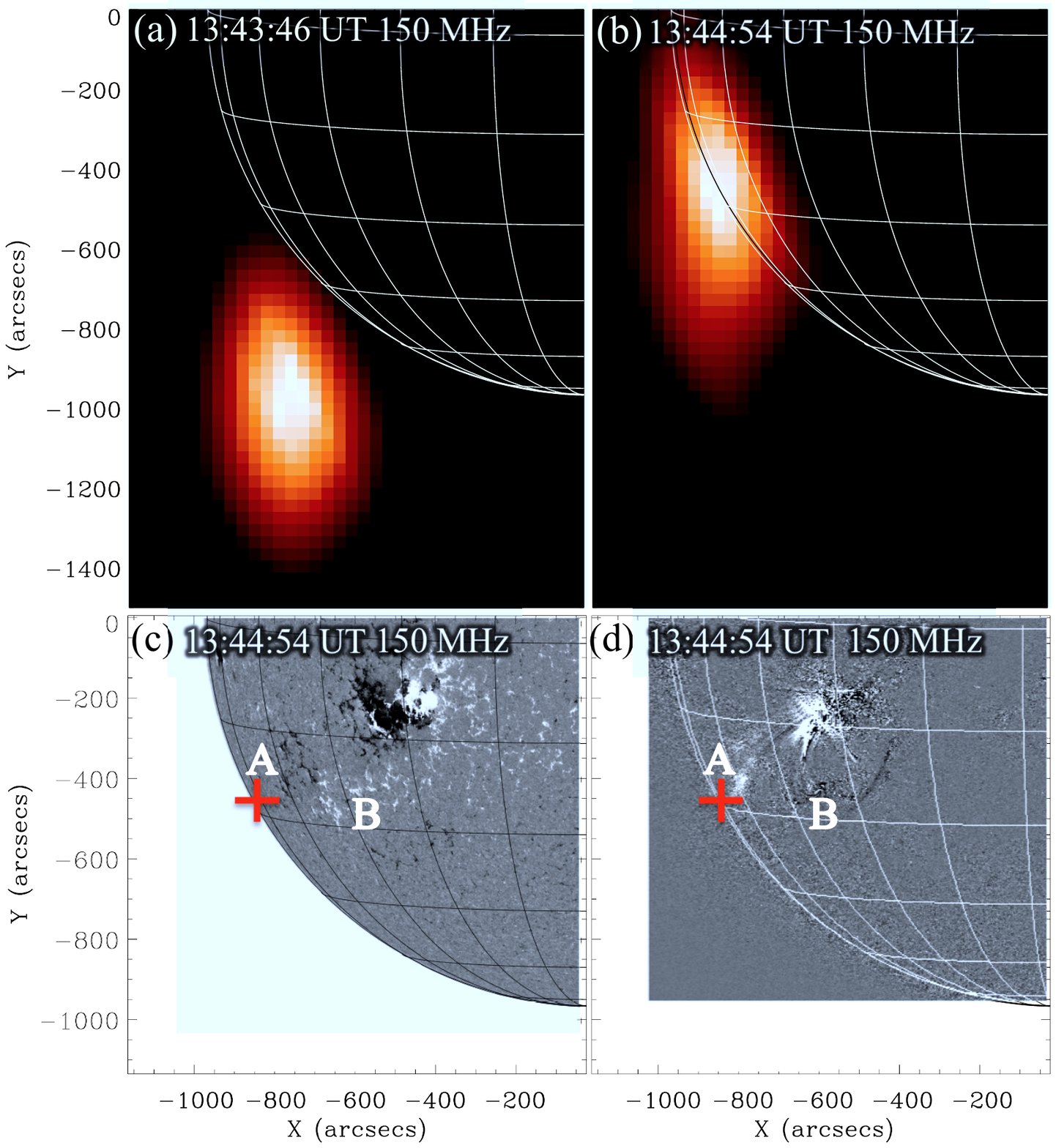}
}
\caption{(a) NRH radio source of the first dm-type~III bursts at 13:43:46 UT; (b) NRH radio source of the first IP-type~III bursts at 13:44:54 UT; (c) the location of the IP-type~III bursts is indicated by a red cross on the $SDO$/HMI magnetogram and, on (d) the $SDO$/AIA 193 \AA\ running difference.}
\label{Figure5}
\end{center}
\end{figure*}

Two successive interplanetary  groups of IP type~III bursts started  respectively at  13:43:57 UT and 13:48:20 UT. The first group took place shortly  after the dm-type~III bursts. The spectrum displayed in Figure~\ref{Figure2} shows that both IP groups are detected at frequencies below $\sim$70~MHz with the DAM spectrograph. The first one was detected by the NRH at 173~MHz and 150~MHz (but the second one was not). Its emission measured at 150 MHz  was not polarized.  The two upper Panels (a) and (b) of Figure~\ref{Figure5} show that the respective locations, at 150~MHz, of the dm and IP type~III bursts are different.  The IP burst location, which is superposed on an HMI magnetogram in Panel (c) is quite close to the noise storm position, above the  A-B coronal loop system (defined in Figure~\ref{Figure3}c and d).


 Moreover, the running-difference image at 193 \AA\ (see Figure~\ref{Figure5}d) shows that these bursts are also concomitant with a sudden brightening, identified in the $SDO$ image at 13:45 UT, which appears on the east side of the eruptive jet, near the B foot of the  A-B coronal region and persists until at least 13:48 UT. Open field lines, identified in the magnetic field line extrapolation are present nearby this brightening. Most of them originate in A. This set of observations leads us to propose that the electron beams, which produce the IP type~III emission, result from the reconnection of these open field lines with the  magnetic field structure of jet.

\subsection{CME and type II radio burst}
\label{Sect:CME_type_II}

The onset of the CME rising loops at 13:44~UT (see Movie 2) was followed, soon after,  by a type~II burst starting at 13:45:59 UT. In this section, we investigate the relationship between the type II progression and the CME evolution.

\begin{figure*}
\begin{center}
\IfFileExists{fastCompil.txt}{
\includegraphics[width=0.9\textwidth,clip]{Figure6}	                
	                        }{
\includegraphics[trim= 1cm 2cm 1cm 2cm,clip=true,width=12cm,angle=90]{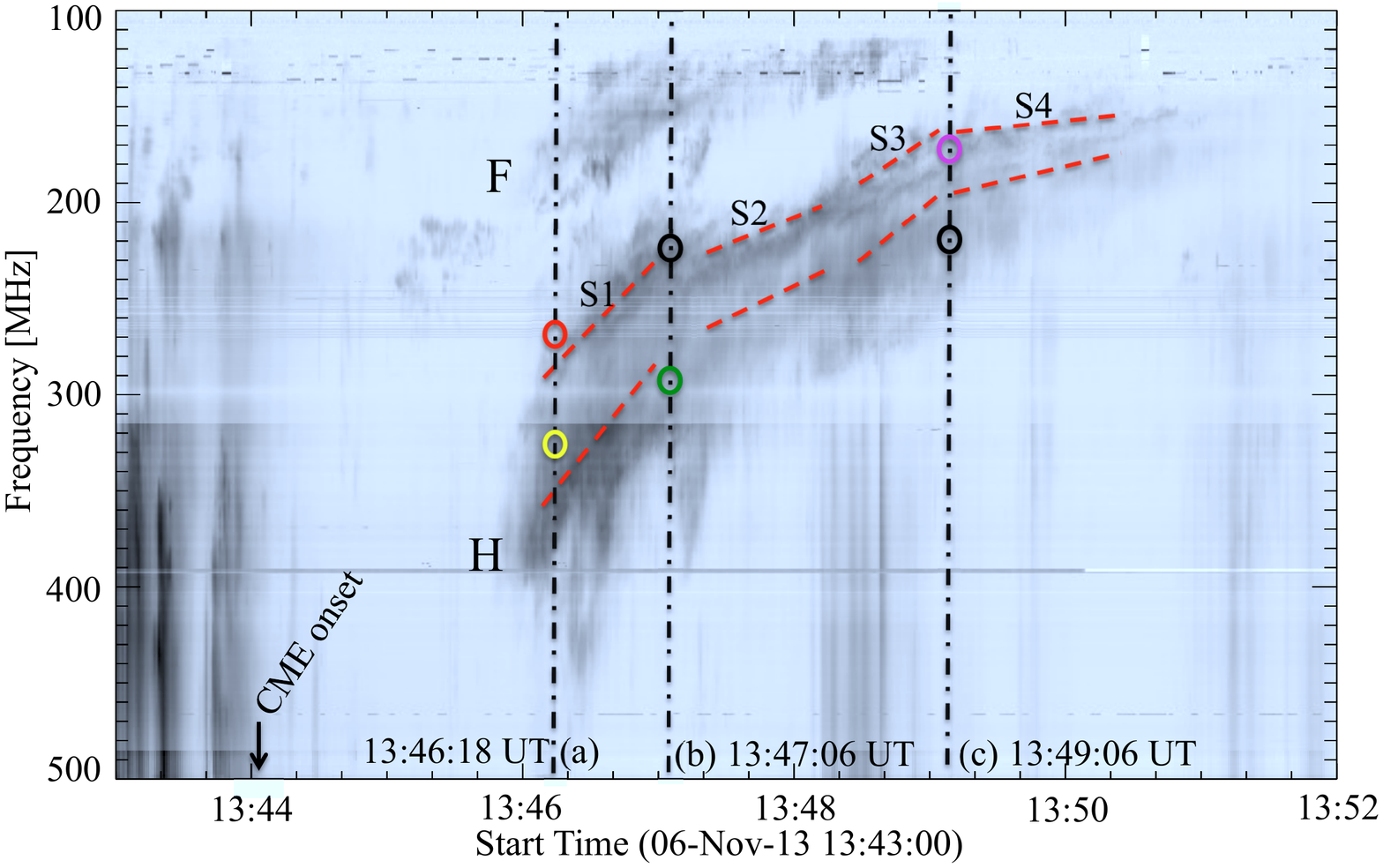}
}
\caption{Dynamic spectrogram of the type~II radio bursts observed with ORFEES and e-Callisto. The burst shows both fundamental (F) and harmonic (H) lanes of emission. The (H) band splitting is indicated with the red dashed lines; this is fragmented in four main segments with different frequency drift rates; each segment is labelled from S1 to S4. The times, where the NRH sources of the type~II splitted bands have been measured, are indicated by black dashed lines (see section 3.3.2); the source  frequencies are indicated by colored circles: the same color symbol as in Figure~\ref{FigureUpDownSources} has been adopted: yellow 327~MHz, red 270~MHz, green 298~MHz, black 228~MHz, and pink 173~MHz. }
\label{Figure6}
\end{center}
\end{figure*}

An expanded view of the type~II burst is presented in Figure~\ref{Figure6}. This burst exhibits the (F) fundamental and (H) harmonic emission bands. The H band is clearly splitted in two parallel lanes. Following the interpretation proposed originally by \citet{Smerd1974}, these two lanes are a consequence of the plasma emission of the upstream and downstream shock regions.


\begin{figure*}
\begin{center}
\IfFileExists{fastCompil.txt}{
\includegraphics[width=0.5\textwidth,clip]{Figure7}	                
	                        }{
\includegraphics[trim=3.2cm 3.1cm 4cm 1.8cm, width=9cm, clip=true,angle=0]{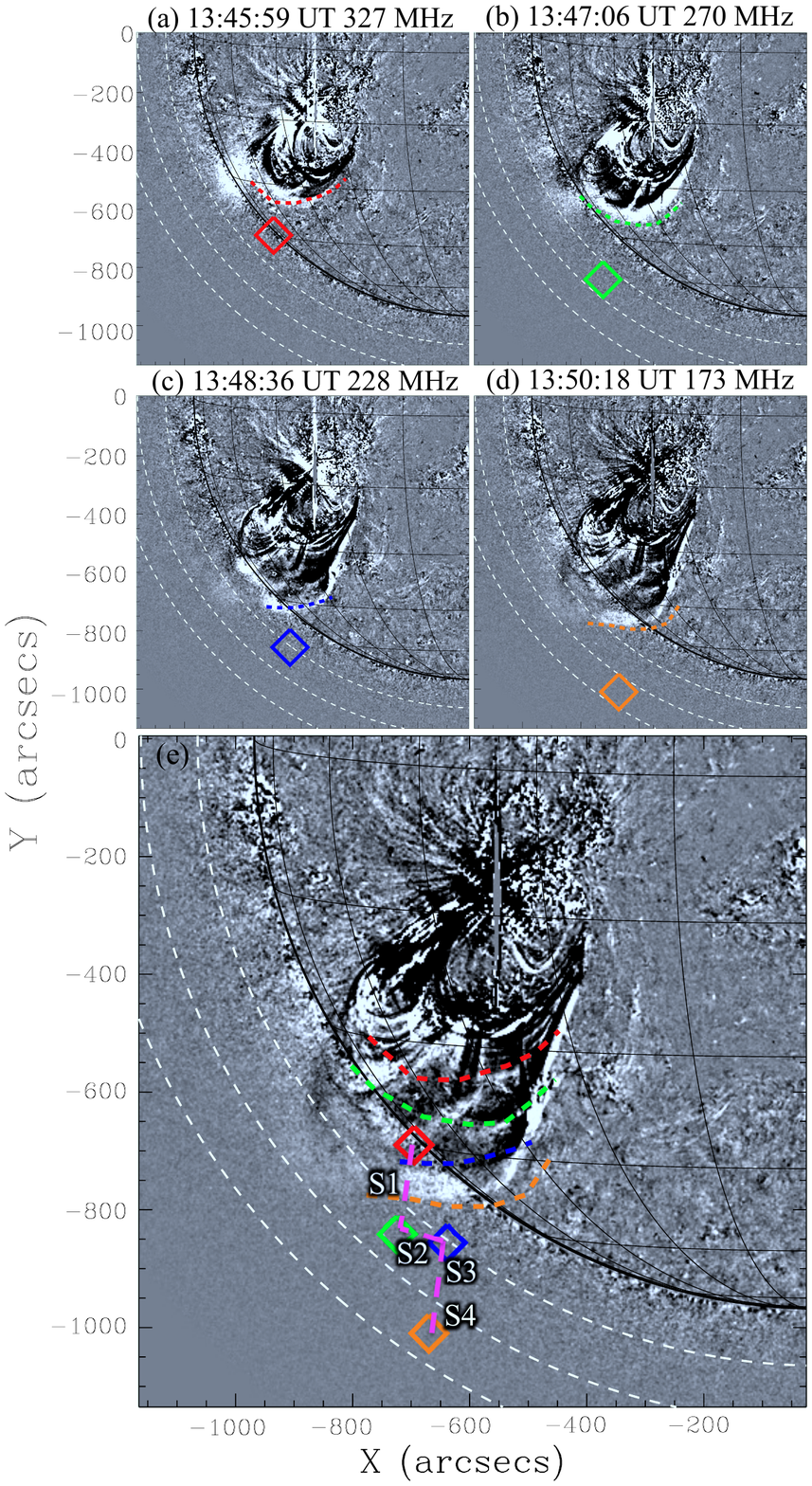}}
\caption{Running difference images of the erupting CME observed with $SDO$/AIA 193 \AA\  with the NRH radio source of the type~II shock superposed. For each Panel, the CME LE is indicated with a dashed line with the same color of the square indicating the NRH source at the same time. (a) Type~II bursts source at 13:45:54 UT; (b) end of first segment at 13:47:06 UT; (c) CME LE with elongated shape in the southern region at 13:49:06 UT; (d) final stage of the type~II burst at 13:50:18 UT, when the type~II burst begins to fade in the spectrum; (e) the superposition of the four locations. The three directions of propagation are indicated with purple dashed lines labelled S1, S2, and S3-S4 corresponding to the segments defined in Figure~\ref{Figure6}).}
\label{Figure7}
\end{center}
\end{figure*}

The H component is fragmented in four main segments, highlighted by  red dashed lines in Figure~\ref{Figure6}. The positions of the upstream type II source, measured by the NRH  at different times and frequencies, are superposed in Figure~\ref{Figure7} on  $SDO$/AIA running difference images at 193~\AA . These images display the progression of the CME. The diamond symbol indicates the position of the upstream radio source, while the location of the CME leading edge (LE) is indicated by a dashed line (see the figure caption for the color code of frequencies).

\begin{figure*}
\begin{center}
\IfFileExists{fastCompil.txt}{
\includegraphics[width=0.7\textwidth,clip]{Figure8}	                
	                        }{
\includegraphics[trim= 0cm 6cm 0cm 0cm, width=14cm,clip=true]{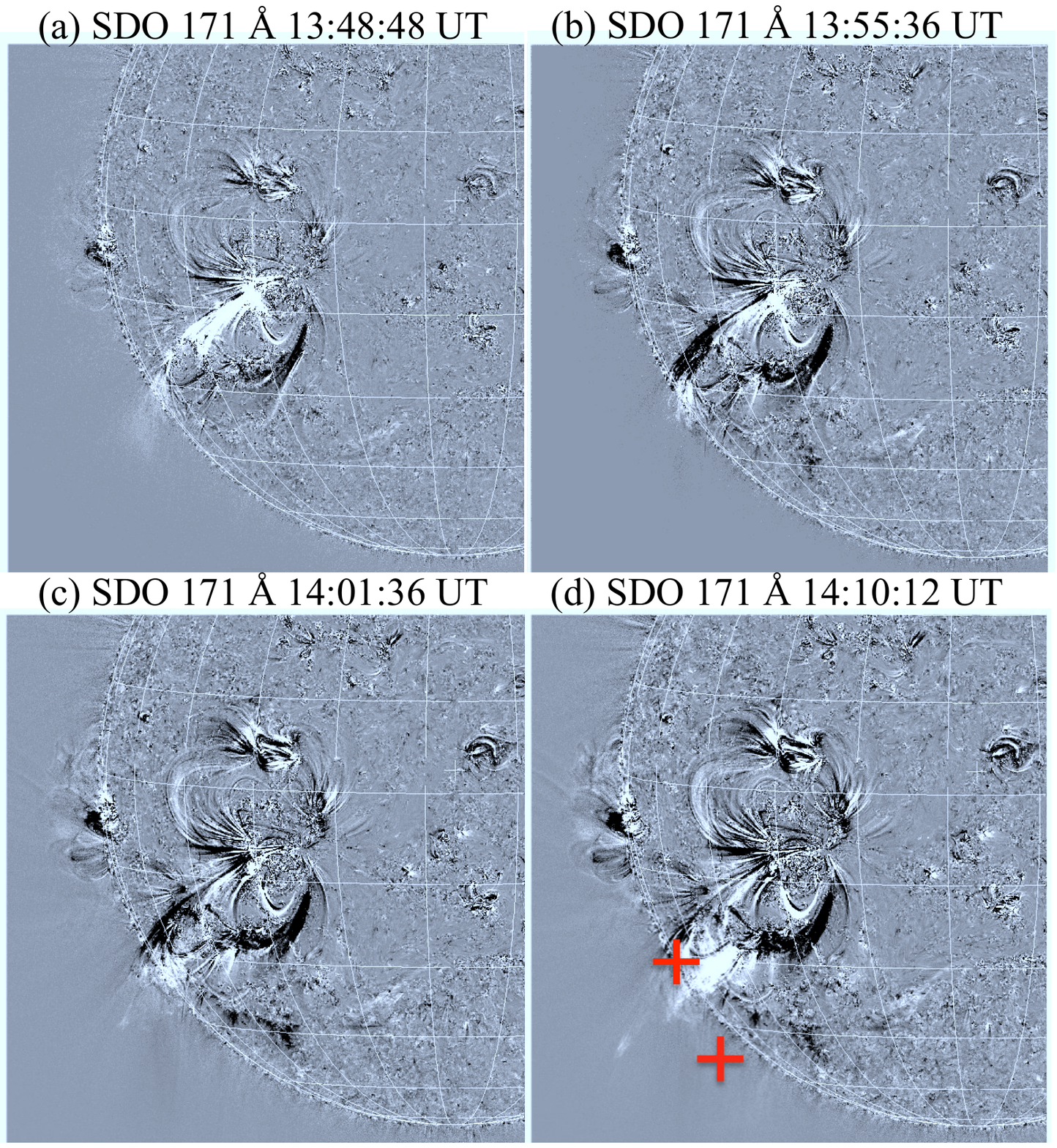}
}
\caption{Four base-difference images of  $SDO$/AIA obtained successively (a) near the end of the westward deviation of the type~II burst (end of S2, see Figure~\ref{Figure7}); (b) near the onset of the  m-type~II burst recorded by the DAM spectrograph (see Figure~\ref{Figure2}, also for (c) and (d)); (c) near the occurrence of the second group of IP type~III bursts; (d) near the occurrence of the third group of IP type~III bursts; the locations of these bursts at 14:10 UT and 14;12 UT are indicated by two crosses.}
\label{Figure8}
\end{center}
\end{figure*}

We note the following sequences in the radio spectrum of the type~II burst:
 
a) The initial position of the upstream type~II burst measured at 13:45:59 UT is located above the LE of the CME (Figure~\ref{Figure7}a). Both the CME LE and the upstream radio sources propagate in the south direction. The distance between the type~II source and the CME LE is increasing; that indicates that the shock source is moving faster than the LE.

b)  At 13:47:06 UT, the dynamic spectra (Figure~\ref{Figure6}) shows an abrupt change in the drift rate of the type~II burst, passing from 1.25 to 0.55~MHz/s  (\textit{end of the first segment}). Between $\sim$~13:47:00 and $\sim$ 13:48:36 UT, the type~II source stops its southward progression, being westward deviated (Figure~\ref{Figure7}b,c and e). During the same period of time, the shape and the orientation of the CME LE are also modified (Figure~\ref{Figure7}c, Movie~4). The CME LE is now southward elongated and slightly westward oriented. We thus conclude that the change in  the frequency drift and in the trajectory of the   type II burst coincides with the change in the orientation of the CME. Moreover, these modifications occur during the  same period as  the bright "like-bridge" which connects the eruptive jet with the  B region.
We thus conclude that the change in the type II burst trajectory and in the orientation of the CME result from their approach from the eruptive jet (see Figure~\ref{Figure3}d) which strongly affects their development. (\textit{end of the second segment}). 

c) After 13:48:20 UT, the type II burst returns again at a drift rate of 1.1~MHz/s (\textit{start of the third segment}). There is another change of its drift rate at $\sim$~13:49 UT (\textit{start of the fourth segment}). It becomes $\sim$ 0.2~MHz/s until $\sim$~13:51:00 UT.  At this time, the type~II emission fades and disappears from the spectrogram, marking the end of \textit{the fourth segment} (Figure~\ref{Figure7}d). 

d) While the CME continues its southward progression, its lateral expansion is limited on one side by the eruptive jet and, on the other side by the neighboring CH (Figure~\ref{Figure8}). Its western edge becomes slightly  westward deviated after 13:49 UT; however, the base-difference images displayed in Figure~\ref{Figure8} show that its lateral expansion seems to be limited by the pressure generated by the neighboring CH and stops its westward progression at $\sim$~13:48:50 UT. 


\textit{The second type II burst}

\begin{figure*}
\begin{center}
\IfFileExists{fastCompil.txt}{
\includegraphics[width=0.7\textwidth,clip]{FigureCME}	                
	                        }{
\includegraphics[trim= 2cm 3cm 2cm 2cm,clip=true,width=11.5cm,angle=0]{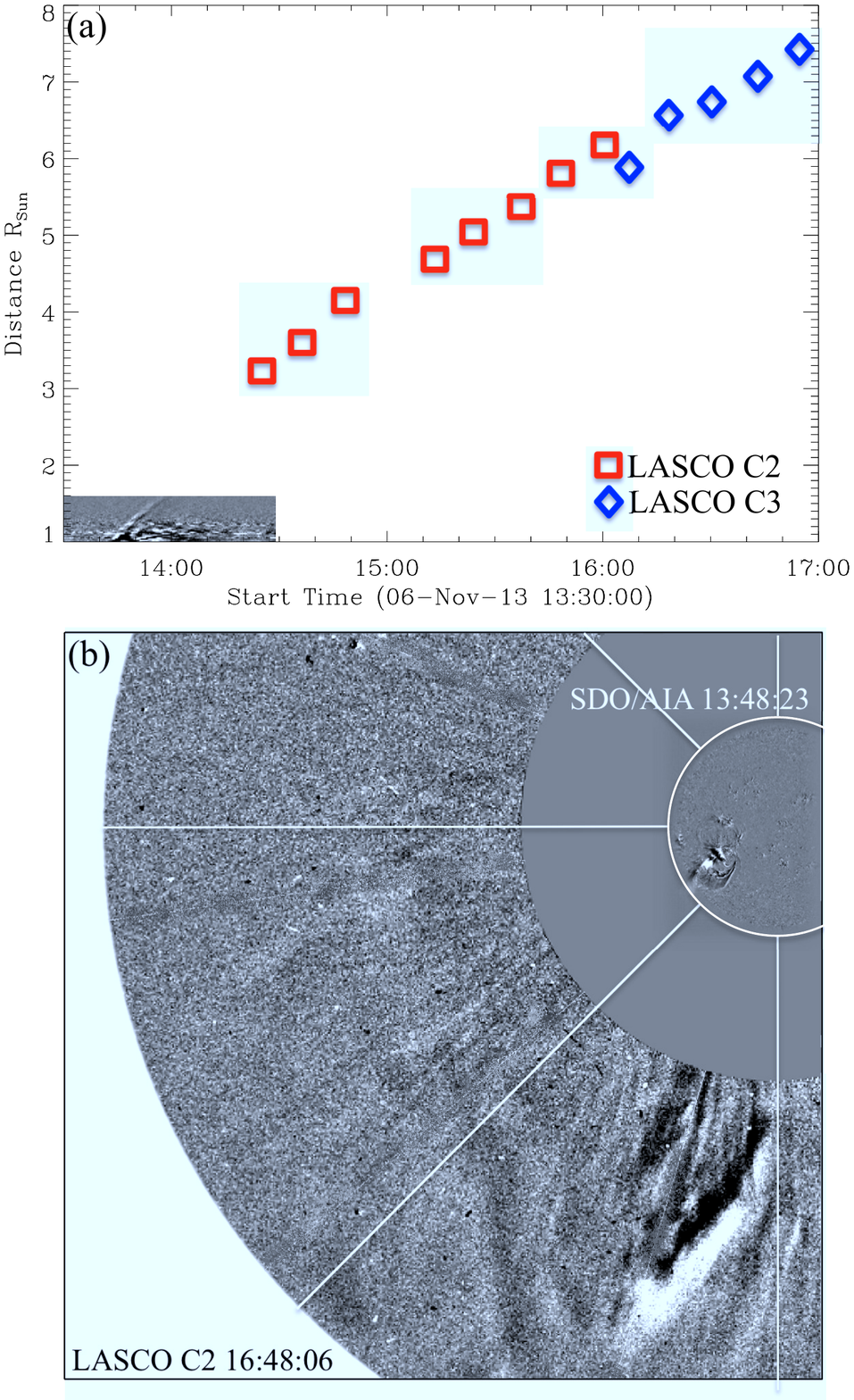}
}
\caption{(a) Height-time plot of the erupting CME measured with SDO/AIA 193 \AA\ and LASCO C2 and C3. (b) Composite plot of running difference image of the CME observed with SDO/AIA 171 \AA\ at  13:48:23 UT and of its later expansion observed with LASCO C2 at 16:48:06 UT.}
\label{FigureCME}
\end{center}
\end{figure*}

 Type~II burst emission reappears at 13:55:30 UT at a frequency below 90~MHz (Figure~\ref{Figure2}), at the same time as the development of a dark feature near the western edge of the CME (see Figure~\ref{Figure8} and Movie 2).   We suggest that the dark feature and the  type~II burst have a common origin attributed to  the pressure exercised by the CH on the CME edge; it was indeed shown that,  the build-up of such  a compression region can be accompanied by compression waves, or shocks detected in EUV and white light images \citep{Vourlidas2003, Yan2006}. This last assumption is consistent with the western compressed shape of the CME edge observed later, at 16:48:06 UT, in LASCO-C2 coronagraph (Figure~\ref{FigureCME}b).
 
e) The m-type~II burst fades abruptly around 14:02 UT, which is the time when the shock reaches the boundary of the southern coronal hole. The spectrogram in Figure~\ref{Figure2} shows the onset of an IP type~III burst, followed a few minutes later by another group of IP type III bursts. The latter were observed by the NRH at 150~MHz; their locations at this frequency are indicated by two crosses reported in Figure~\ref{Figure8}d. These bursts are  probably due to the interaction between the CME LE (or the jet) and the open magnetic field lines in the polar region.  

The progression of the SDO CME is later observed by LASCO C2-C3. A distance-time plot of the erupting CME is shown in Figure~\ref{FigureCME}a. This plot was obtained using the running difference images of the SDO/AIA 193 \AA\ for the range 1-1.6 R$_{\odot}$ and of SOHO/LASCO C2 and C3 for the range 3-8 R$_{\odot}$. A composite running difference image of the CME observed with SDO/AIA at 13:48:23 UT, and at its later expansion stage with LASCO C2 at 16:48:06 UT is shown in Figure~\ref{FigureCME}b; the western edge of the CME appears to be compressed by the interaction of the CH at the western side of the AR.


 The initial velocity of the CME, as measured with LASCO C2 at 3 R$_{\odot}$, is of $\sim$400 km s$^{-1}$ and its final velocity at 8 R$_{\odot}$, observed by LASCO C3, is $\sim$280 km s$^{-1}$, with an acceleration of -13.8 m s$^{-2}$.


\section{Shock and ambient medium characteristics}
\label{Sect:Shock}

In this section, we describe the shock properties, relating its kinematics to the kinematics of its driver, and to the ambient medium characteristics. For that we use the well observed splitting of the type~II harmonic emission in two parallel lanes (Figure~\ref{Figure6}) which correspond to the plasma emission of the upstream and downstream shock regions.

\subsection{The upstream and downstream shock regions}
\label{Sect:up_downstream}

\begin{figure*}
\begin{center}
\IfFileExists{fastCompil.txt}{
\includegraphics[width=\textwidth,clip]{FigureUpDownSources}	                
	                        }{
\includegraphics[trim= 4cm 0cm 3cm 0.7cm,clip=true,width=8.5cm,angle=90]{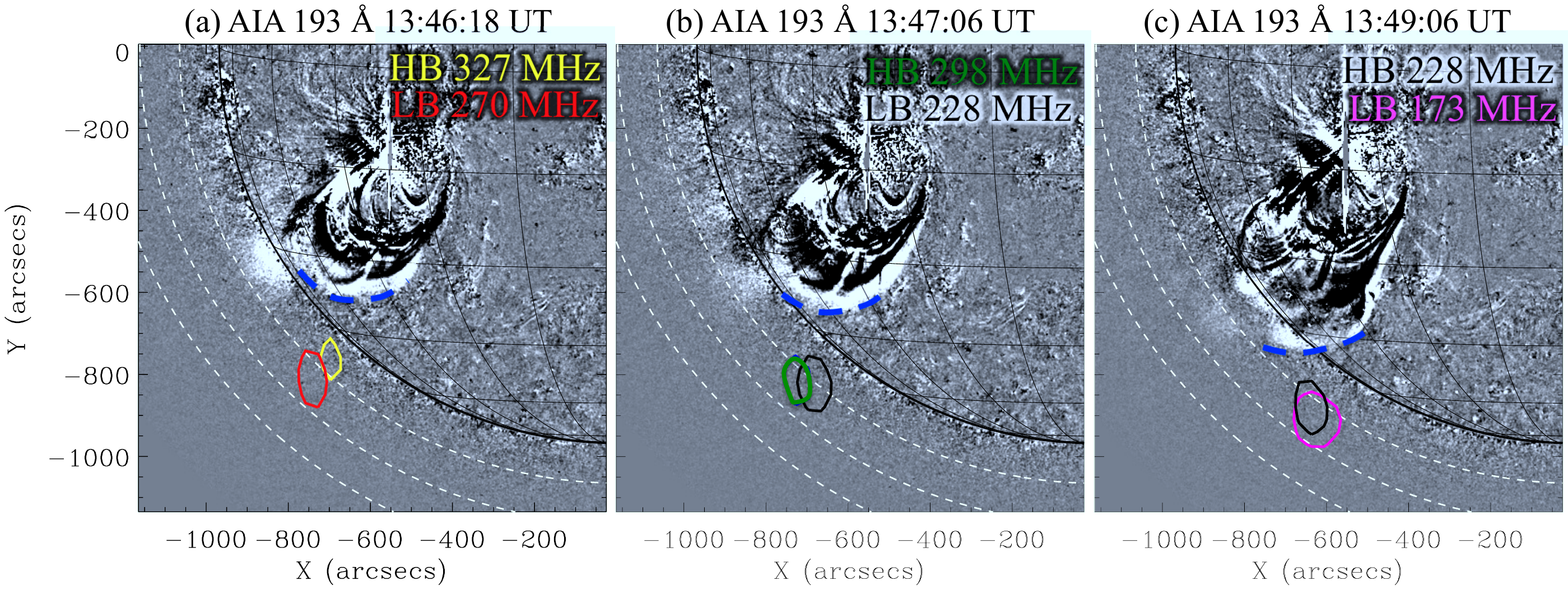}
}
\caption{Location of the resolved band splitting source superposed to the $SDO$/AIA running difference images. The contours (90\% of the NRH peak flux) measured with the NRH, show the up-stream and down-stream shock emission, (a) for the high band (HB) at 327~MHz in yellow and for the low band (LB) at 270~MHz in red at 13:46:18 UT, (b) for the HB at 298~MHz in green and LB at 228 in black at 13:47:06; (c) and for the LB in pink at 173~MHz 13:49:06 UT. The selected points are indicated with the same color code as in the dynamic spectrograph (Figure~\ref{Figure6}). The location of the CME LE is indicated with a blue dashed line.}
\label{FigureUpDownSources}
\end{center}
\label{}
\end{figure*}

 The NRH observations  obtained at the 9 frequencies quoted above allow us to determine the respective location of the  upstream and downstream regions. Figure~\ref{FigureUpDownSources} shows, at three different times, the respective positions of these  two components and of the CME LE, which are superposed on  $SDO$/AIA running difference images at 193~\AA . The CME LE is highlighted with a blue dashed line. Each NRH source position is indicated by a contour (at 90\% of the peak flux), the color  referring to the selected frequency (see the figure caption). For the three chosen times, 13:46:18 UT, 13:47:06 UT and 13:49:06 UT, it was possible to locate simultaneously the upstream and downstream regions. The three couples of selected points are reported in the spectrum in Figure~\ref{Figure6}, using the same color code as in Figure~\ref{FigureUpDownSources}. We note that the orientation of  the two sources in Panel (b) is different from those of Panels a and c. This observation  is consistent with the sudden change of the type~II orientation described in Section~\ref{Sect:CME_type_II} (during the segment 2, Figure~\ref{Figure7}).

\begin{figure*}
\begin{center}
\IfFileExists{fastCompil.txt}{
\includegraphics[width=0.8\textwidth,clip]{FigureDistance-Time}	                
	                        }{
\includegraphics[trim= 1cm 3.1cm 1cm 3.3cm,clip=true,width=13cm,angle=0]{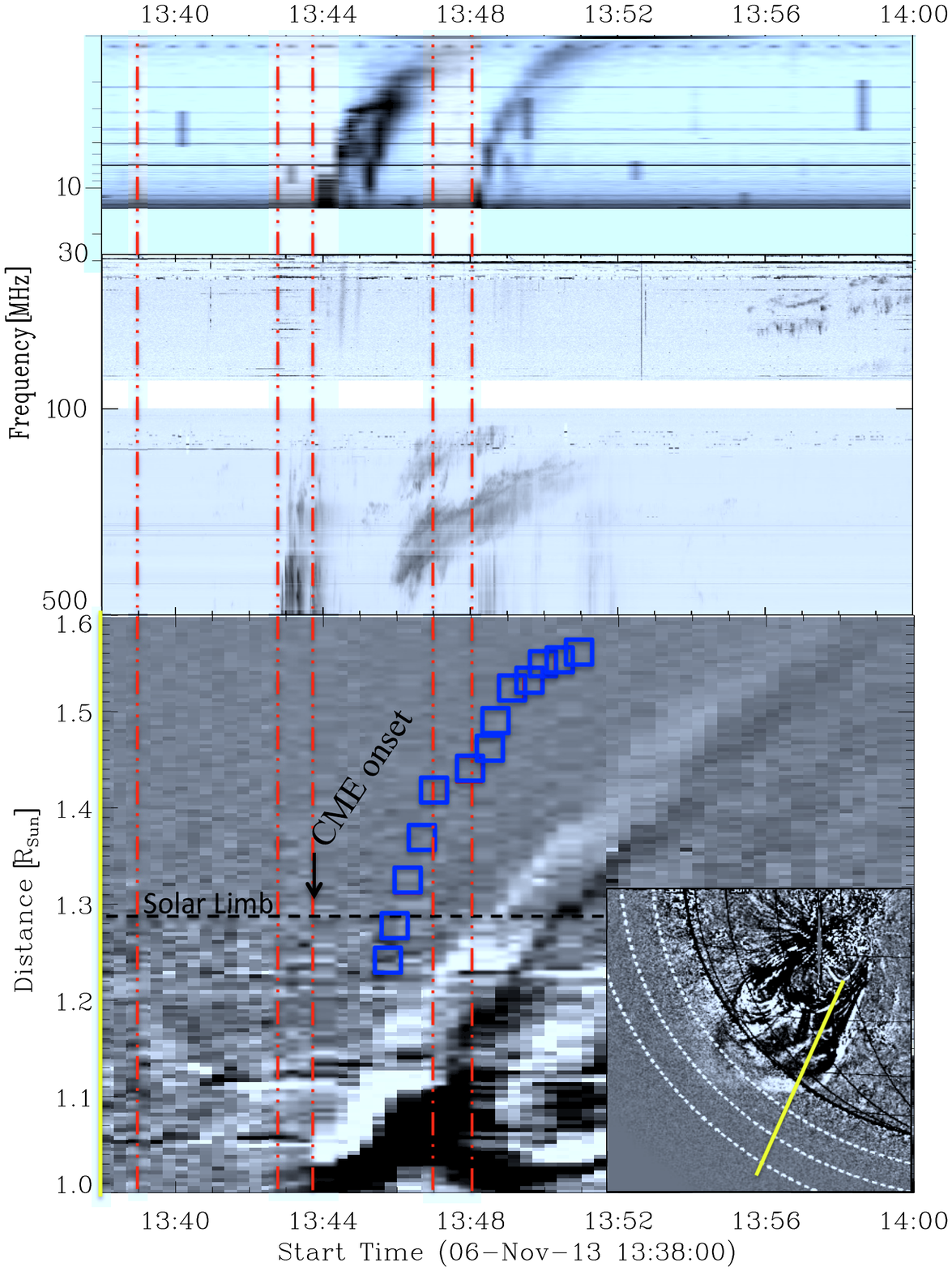}
}
\caption {Dynamic spectrogram (top) and projected distance--time plot of the erupting CME and of the shock (bottom). The trace of the CME distance-time is indicated in yellow in the inset at the bottom right. The NRH shock positions of the up-stream region are indicated in blue. The red vertical lines indicate successively the flare onset time, the onset and end of the first dm-type~III bursts, and the time period when the type~II drift rate shows an abrupt decrease (\textit{second segment}; see Figure~\ref{Figure6}).}
\label{FigureDistance-Time}
\end{center}
\end{figure*}

The timing of the events is shown in the top panel of Figure~\ref{FigureDistance-Time} with
the radio dynamic spectrum.  The red dashed lines correspond, successively, to the starting time of the flare, 13:39 UT, the starting and end time of the first group of dm-type~III, and to the time period of the second segment (when the type~II drift rate is lower).  In the bottom panel of Figure~\ref{FigureDistance-Time} we compare  the \textit{projected} distance-time plots of the CME  LE and of the NRH sources for the H-low band of the type~II burst (blue squares). 


This distance is measured along the direction indicated in yellow in the inset. As already seen in Figure~\ref{FigureUpDownSources}, the upstream position is located in front of the CME LE and increases its separation from the LE, as the former travels faster.

\subsection{Speed comparison}
\label{Sect:Speed}


\begin{figure*}
\begin{center}
\IfFileExists{fastCompil.txt}{
\includegraphics[width=\textwidth,clip]{FigureVA_MA_B}	                
	                        }{
\includegraphics[trim= 3cm 3cm 2cm 3.6cm,clip=true,width=12.5cm,angle=90]{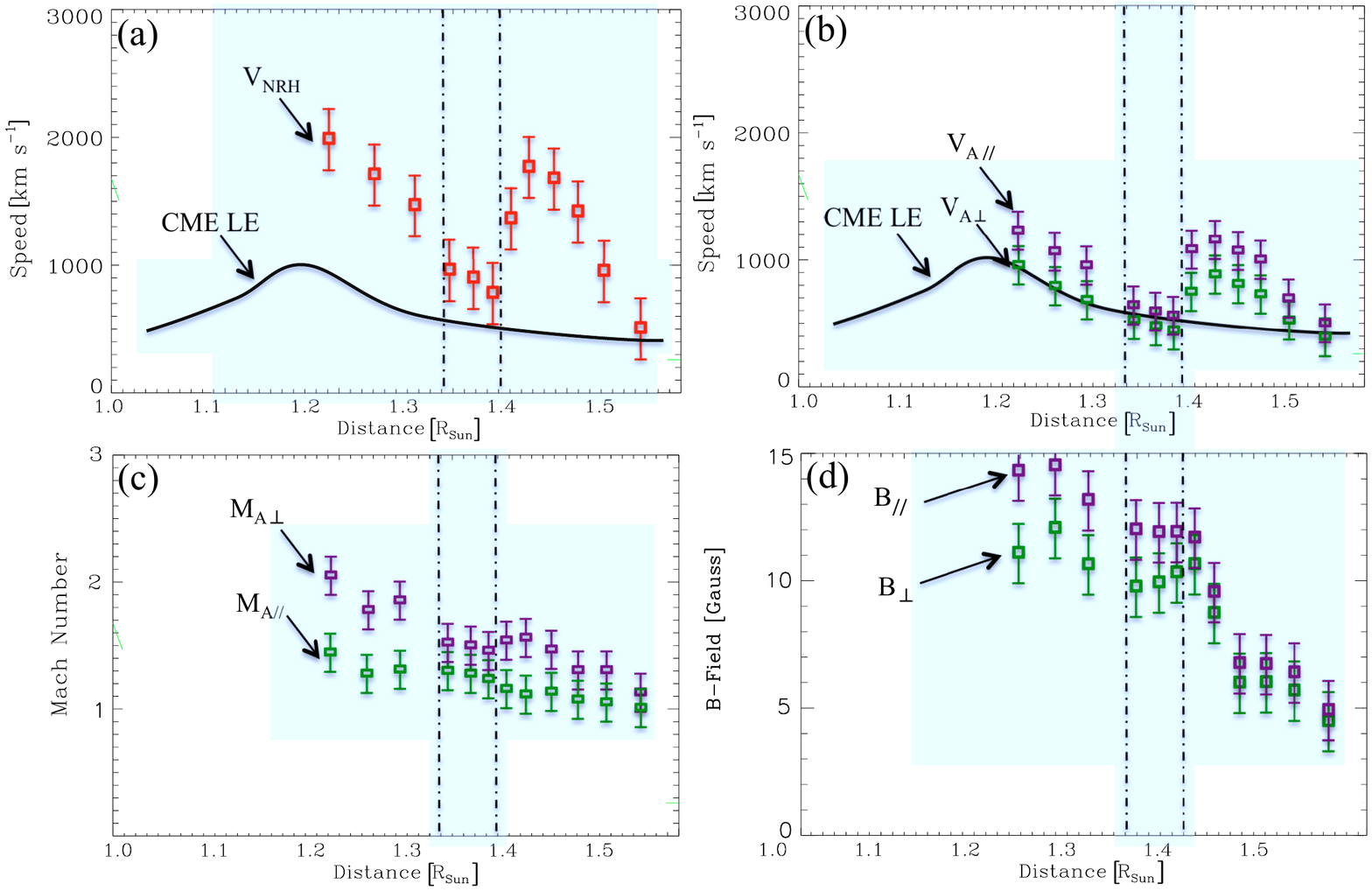}
}
\caption{Speed comparison of the CME LE with the shock and the ambient medium characteristics. (a) The CME LE speed is indicated with a black line, while the speed of the NRH source is indicated with red squares. (b) The Alfv\'en speeds  for the perpendicular and parallel cases are respectively indicated by green and  violet colors. (c) The shock Mach number for both perpendicular and parallel cases. (d) The estimated B-field for the two cases. The vertical dashed lines indicate the start and end of segment 2 when the frequency drift is interrupted (See Figure~\ref{Figure6}).}
\label{FigureVA_MA_B}
\end{center}
\end{figure*}

Figure~\ref{FigureVA_MA_B}a shows a comparison between the  projected speeds of the CME-LE  (black line) and of the type~II NRH sources.  A 2D Gaussian fit was applied to obtain the  location of each radio source at a given time frequency,  i.e.  at a given density.   The  uncertainty  on the NRH speed  estimation depends exclusively  on the uncertainty on the source location; the error bars are plotted at 2$\sigma$. Speeds are here projected speeds. As we do not know the angle between the CME or the type II burst and the plan of the sky, the CME, shock and Alfv\'en speeds given below may be underestimated by $ \sim$ 30 $\%$. As a consequence, the magnetic fields may be underestimated by the same amount. The projection effect does not affect comparison between the CME and Alfv\'en speeds, because the two speeds are likely to have the same angle with the plan of sky.
 
 The starting speed of the CME is $\sim$~300 km~s$^{-1}$  and reaches a maximum value of $\sim$~1000 km~s$^{-1}$ at $\sim$13:45:50 UT. The speed of the shock up-stream region is initially of $\sim$~2000 km~s$^{-1}$ (\textit{first segment} as defined in Figure~\ref{Figure6}); it then progressively decreases to  $\sim$~800~km~s$^{-1}$ (\textit{second segment}), before increasing again to $\sim$1800 km~s$^{-1}$ (\textit{third segment}) while finally  reaching the low value $\sim$~400~km~s$^{-1}$ (\textit{fourth segment}) when the type~II emission fades out. 


\begin{figure*}
\begin{center}
\IfFileExists{fastCompil.txt}{
\includegraphics[width=0.9\textwidth,clip]{Figure6}	                
	                        }{
\includegraphics[trim= 2cm 3.5cm 2cm 3.5cm,clip=true,width=10cm,angle=90]{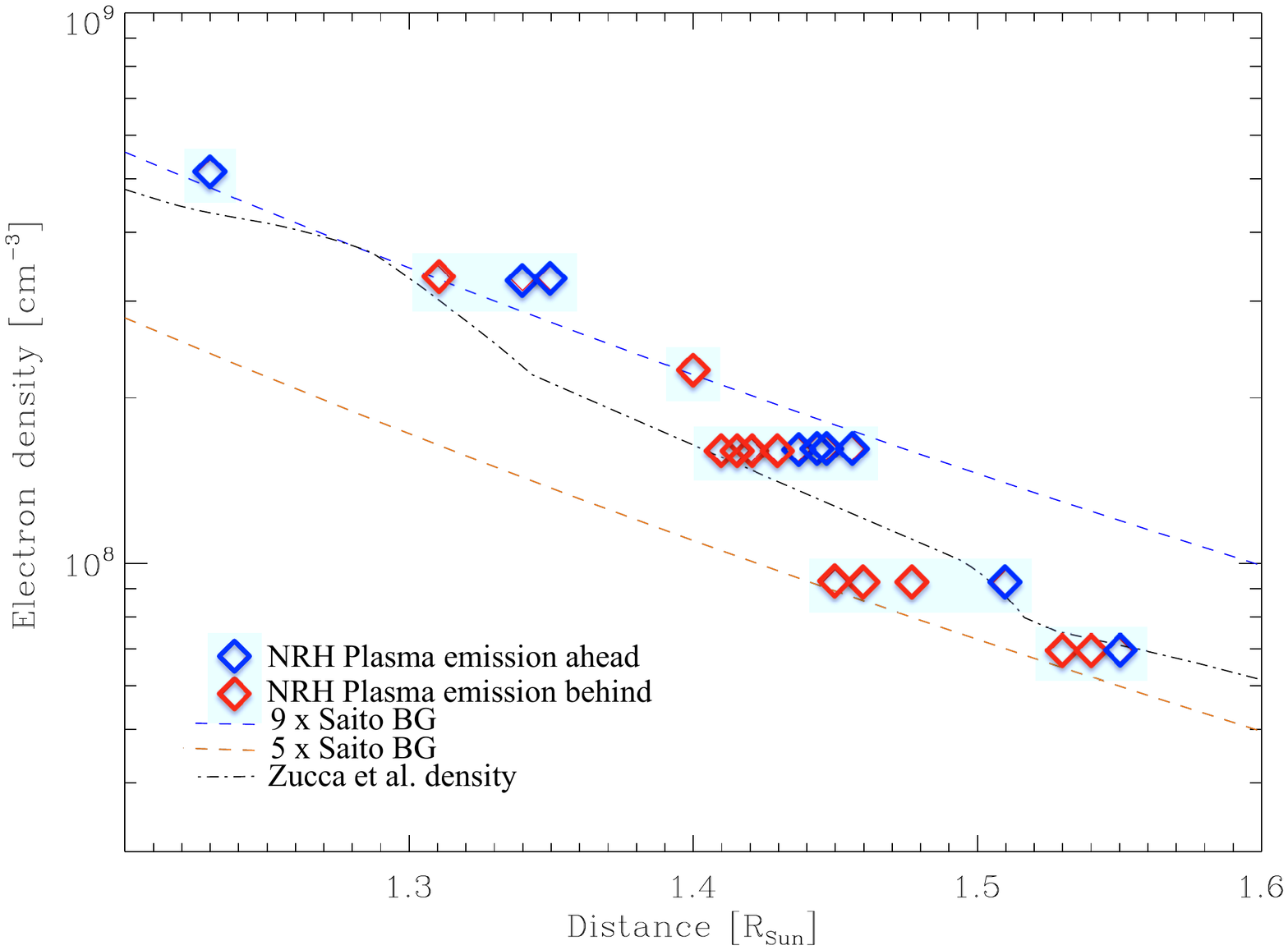}
}
\caption{Electron density assuming harmonic plasma emission of the type~II radio burst source observed with NRH at 408, 327, 270, 228, 173, 150~MHz  against its projected distance. Plotted for comparison the 5xfold and 9xfold Saito background corona model \citep{Saito1977}, and the density estimation using the six EUV filters on $SDO$/AIA \citep{Zucca2014}. }
\label{Figure_extra}
\end{center}
\end{figure*}

To establish a comparison between the speeds of the CME LE and of the type~II shock with the ambient Alfv\'en speed, $V_A$, we applied the procedure developed  by \citet{Vrsnak2002} for both the quasi perpendicular and the parallel shocks. In the present case, the direction of propagation of the shock, obtained from the locations of the NRH source, is indicated in Figure~\ref{Figure7}e by the three purple segments labeled by the numbers  S1-S2, and S3-4 (the direction is the same for the segments 3 and 4).  During the segment S1, the shock propagates through closed loops (see Figure~\ref{Figure3}a) and the normal to  the shock is quasi perpendicular to the direction of the magnetic field. In the cases of S2 and S3-S4, the coronal extrapolation is too complex to establish  a definitive estimate. Thus, we present here the results  obtained for both quasi  parallel and perpendicular cases.

 The procedure developed by \citet{Vrsnak2002} shows that one can obtain the Mach number $M_A$ then the Alfv\'en speed,  if the compression ratio of the shock is determined. This ratio, \textit{X}, is given by $X=(f_u/f_l)^2$, where $f_u$ and $f_l$ are the frequencies of the upper and lower bands respectively. For a low plasma parameter $\beta\rightarrow 0$, the Mach number can be written as $M_A\approx(X(X+5)/2(4-X))^{0.5}$ for a quasi perpendicular shock. As the speed of the shock, $V_{shock}$ is known from the NRH source, the ambient Alfv\'en speed can then be calculated from $V_A=V_{shock}/M_A$. The same procedure can be used for the case of a quasi-parallel shock, in this case the Mach number is given by $M_A=(X)^{0.5}$.

The calculated Alfv\'en speeds for both perpendicular and parallel cases are plotted on Figure~\ref{FigureVA_MA_B}b, green squares for the perpendicular case and violet for the parallel case, together with the CME LE speed. The CME LE reaches a super-Alfv\'enic speed, for the perpendicular case, at a projected distance of $\sim$1.2~R$_{\odot}$ corresponding to the onset time of the type~II shock at $\sim$13:45:59 UT. This agrees with the hypothesis that the shock wave is generated by the CME LE in a piston-driven scenario \citep{Zimovets2012}. 

\subsection{Others ambient medium characteristics}
\label{Sect:Others}

The Mach number obtained from the compression ratio is plotted in Figure~\ref{FigureVA_MA_B}c. The initial value for the perpendicular case of $\approx 2.1$, decreases to the final Mach number of $\approx1.2$; this is indicative of the shock speed approaching the local Alfv\'en speed, at this time the type~II burst fades and disappears from the dynamic spectra at 13:51 UT. Similar but slightly lower values of the Mach number are found for the parallel shock case.

In conclusion, this analysis confirms that the approximation of a quasi perpendicular approach for direction 1 looks reasonable while, no firm conclusion on the type of shock can be given for segments S2 and S3-S4. Still the results of Figure~\ref{FigureVA_MA_B} for those segments are comparable between parallel and perpendicular cases.

The strength of the local magnetic field B can be also estimated from the Alfv\'en speed using the equation: $B=V_A(\mu\rho)^{-0.5}$, where the coronal plasma density is approximately given by $\rho\approx m_p n_e$. The magnetic field values are shown Figure~\ref{FigureVA_MA_B}d for the perpendicular and parallel approaches. We found values for the B-field ranging from 10 Gauss at the beginning of the shock at 13:45:59 UT to 5 Gauss at 1.5 R$_{\odot}$  when the type~II burst fades at 13:51 UT.

\subsection{Electron density at the shock location}
The NRH observations at different frequencies allow us to estimate the electron density at the shock location. The only assumption at this stage is that the type~II emission is due to harmonic plasma emission. The electron density can be then estimated directly from the emitting frequency $f_p\approx18000 \sqrt{n_e}$. The electron density obtained from the NRH frequency bands at 408, 327, 270, 228, 173 and 150~MHz is plotted against the projected distance of the NRH source in Figure~\ref{Figure_extra}. The emission of the upstream component in the low-frequency band (ahead) is plotted with blue diamonds, while the emission of the downstream component (behind) is plotted with red diamonds.
This density is compared with the coronal background electron density using a 5xfold (brown dashed lines) and 9xfold Saito (blue dashed lines) density model \citep{Saito1977}. The position of the radio source for projected distances below $\sim$1.4~R$_{\odot}$ is comparable with a dense 9xfold Saito due to the presence of the active region, while at projected distances over $\sim$1.4~R$_{\odot}$ the density drops and it is comparable with a 5xfold Saito. This is due to the change in direction of the radio burst source as indicated in Figure~\ref{Figure7}e. The density `jump' from a 9xSaito to a 5xSaito of the radio source is evident from the data (see Figure~\ref{FigureVA_MA_B}). This change in density is due to the transition from the dense plasma in the closed magnetic field topology over the active region to the less dense plasma of the other neighboring closed loops structure. This change in the magnetic field topology and electron density may also indicate the reason of the sudden change of direction of the radio source.

An actual estimation of the electron density along the path of the radio source (yellow line on the inset of Figure~\ref{FigureDistance-Time} can be obtained from the 2-dimensional maps using the method described in \citet{Zucca2014}. This method calculates the electron density from emission measure (EM) maps derived using the $SDO$/AIA's six coronal filters and the method of  \citet{Aschwanden2013}. The plasma electron density can be calculated from the EM by estimating an effective path length of the emitting plasma along the LOS (see \citet{Zucca2014} for details). The electron density calculated using this method is plotted in Figure~\ref{Figure_extra} with a black dashed line. The density profile is compatible with the mean density profile obtained  from the NRH and spectral observations.

\section{Summary and Discussion}
\label{Sect:Discussion}

\subsection{Jet and CME: the joint evolution}

In the present paper, we have presented an unusual event in which an eruptive jet is involved in the onset of a CME and, then, accompanies its development. First detected in EUV, this event appears as a simple loop system rising in the corona, while it is later identified as a CME when observed, in white light, by SOHO. No EUV plasmoid was detected behind the edge of the CME.

The aim of this paper was to understand:  i) the role of the eruptive jet and of the ambient medium for setting up  this CME; ii) the nature of the two shocks associated with two successive type II bursts,  which occur during  the CME progression. Our main findings are summarized below.

A)
 The eruption of the jet marks the beginning of the event; this jet originates from a coronal null point above a  positive parasitic polarity embedded inside the trailing negative part of the Active Region (AR). The initiation phase occurs when  a sudden brightening appears at its base and extends rapidly toward its western neighboring loops.  These large scale loops, which connect the two main polarities of the AR, start to shine  along their eastern leg. These observations are indicative of the beginning of a destabilization process of the loops caused by their magnetic interaction with the jet structure.

B)
The subsequent eruptive EUV and radio manifestations, preceding or accompanying the onset of the CME, occur in the vicinity of the region where the CME LE is formed. The idea of a magnetic reconnection process between the outer part of the jet and the ambient medium, is confirmed by the presence of radio type III bursts. 

 	a) The first dm type III burst group, which  is followed by two other ones, coincides with the appearance of a bright narrow EUV source located above a region of positive polarity. This source persists during several minutes, and subsequent weaker sources are observed in the same region. These observations suggest that the sources, from which the electron beams responsible for the type III radio bursts originate, result from a magnetic reconnection between the western side of the eruptive jet magnetic field and the field lines anchored in this region of positive polarity. This interpretation is consistent with i) the sudden appearance, between these two regions, of a bright bridge, particularly well observed by SDO at high temperature; ii) the positive polarization of the radio emission; iii) the trajectory of the electron beams (Note that the magnetic field lines of positive polarity are not detected by PFFS). 

	 The ascending motion of the CME starts soon after the occurrence of the first group of type III bursts. 

	b) During these dm bursts, two groups of interplanetary type III bursts are also detected. The electron beams producing these bursts  result from the interaction between the eastern side of the jet and the open field lines originating from a region of negative polarity.

C)
The different following observations seem to stress the role of the erupting jet during the CME progression:

	a) Its motion is  followed by the onset of the first type II burst. The progressions of the CME and of the burst appear to be closely connected: they follow the same direction and the type-II burst sources are located above the front edge of the CME. Moreover, approximately 1 or 2 minutes later, the source of the type II burst stops its southward motion and becomes westward oriented, while  the CME leading edge becomes also slightly westward oriented.  We attribute this effect to the encounter of the eastern edge of the CME with the eruptive jet. Let us further remark that: i) its eastern edge becomes no longer discernible from the jet after their encounter; ii) the CME leading edge appears to be split into two parts, its eastern part corresponding in fact to the western branch of the jet, now curved and surrounding the CME. These facts are compatible with the shape of the CME as observed later.

	b) While the CME continues its southward progression, its lateral expansion is limited, on one side by the presence of the eruptive jet, and on the other side by the pressure generated by the coronal hole. This interaction is possibly the cause of the  second type II burst which, during the same period, appears at decameter wavelengths. We note that, 2 hours later, the expansion of the CME, when observed by LASCO/C2, has remained the same.

	c) The last IP bursts, which appear approximately at the time of the approach of the CME with the south pole, probably originate from a magnetic interaction between the CME, or the jet itself, with the open field lines of the polar region.
 
\subsection{First shock and ambient medium characteristics}

In this study, the CME LE and the type II burst kinematics were compared with ambient coronal characteristics such as the Alfv\'en speed and the B-field, in order to understand the origin of the shock and its progression. These properties were calculated without assuming any model for the coronal density and they were derived from the shock compression ratio; the latter was obtained from the type II split lanes, using a method described in \citet{Vrsnak2002}.  

The CME LE showed a fast initial acceleration, and already reached a super-Alfv\'enic speed. This was subsequently followed by the production of a type~II burst with emission lanes split in two bands. The type~II burst also presented a fast initial acceleration leading to a speed faster than the CME LE, so that they progressively separate one from the other.

A shock can be a blast wave, in which the energy is supplied by a pressure pulse, or it can be driven by a CME, either in a piston-type or in a bow shock scenario \citep{Vrsnak2005}. In the case of a piston shock geometry, the shock moves faster than its driving piston and the medium is confined, since it is not able to stream around the CME \citep{Vrsnak2005, Warmuth2007}.  In our event, as recalled in point C-b above, the  lateral expansion of the CME is limited, on one side by the presence of the eruptive jet, and on the other side  by the pressure generated by the coronal hole. This confinement,  together with the shock propagating faster than the CME LE, strongly suggest that the shock has been driven by the CME in a piston-driven mechanism. Another observation is in line with this interpretation: the type II burst sources are  located in front of the CME LE and undergo the same change as the CME in the propagating direction. 

The radio observational coverage by the NRH allowed us to resolve the location of the split bands of the type II burts. We found that the two components were located ahead of the CME LE and that the higher frequency lane was positioned behind the lower frequency band.
This is in agreement with the \citet{Smerd1974} interpretation of the splitting lane emission. In our scenario, the hypothetic shock wave, probably to faint to be detected in EUV, is located between the low and hi band position of the splitted lanes. \citet{Bain2012} and \citet{Zimovets2012} arrived to a similar conclusion, in the study of another dm-metric type II burst which was also imaged by the NRH. 

\section{Conclusions}
\label{Sect:Conclusions}
We have presented the formation and development of an unusual CME as described in Section 4 and summarized below. The CME resulted from the interaction of an eruptive jet with the surrounding medium. The key points are the overall magnetic structure of the ambient medium and the relative position of the jet in this environment. To our knowledge, it is the first time that  such an  event has been analyzed in some depth. A cluster of eruptive EUV and radio observations, stress the predominant role played by the  eruptive jet in the history of this CME:

	First detected in EUV, this event appears as a simple loop system rising in the corona.  These loops start to be destabilized by their magnetic interaction with the jet. This early development of the CME does not show the signatures that could be expected from previous observations (see introduction).

	 Then, a destabilization process of the loops is caused by  magnetic reconnection  between the outer part of the jet and the ambient magnetic field. This process occurs in the vicinity of the region where the CME LE is formed and when the CME speed is strongly increasing. This is also near this time that the onset of the first type III burst is observed. This is reminiscent of the break-out model \citep{Antiochos1999} but with reconnection between closed and open magnetic field. The progression of this  CME is later observed in white light, up to a distance of 8 solar radii.

	Two type II bursts were detected. A distinct origin is identified for the two successive shocks, both associated with the CME development.  One of the primary finding of this study is related to the first type II burst for which the joint spectral and imaging observations allowed us:

	- To identify step by step the origin of the spectral fragmentation, in relationship with the CME evolution; 

	- To obtain at each step, without introducing an electronic density model or a MHD simulation, the upstream plasma density, the  Alfv\'enic Mach number for the shock and the magnetic strength.

	The jet and/or CME are at the origin of interplanetary radio type III bursts; these bursts  reveal the injection, in the interplanetary medium, of electron beams along different directions.

To conclude, we would like to illustrate, on two specific points,  how the data analysis has benefited from particularly favorable conditions: i) Though, the event originated on the solar disk, it was observed by the SOHO/LASCO coronagraph. It allowed us to confirm that this event was a real CME; ii) the polarization measurements of the radio type III bursts was determinant to identify the origin of the dm type three bursts and also showed that electron beams escape along magnetic field lines that were not present in PFSS extrapolation.

\textbf{Acknowledgements}

P. Zucca is supported by a TCD Innovation Bursary and acknowledges the CNES for the financial support he received during his stay in LESIA, Meudon Observatory.  We thank E. Pariat  for constructive discussions. We would like to thank the referee for the valuable comments and suggestions. We are also grateful to the SDO team for his open data policy. The SOHO LASCO data used here are produced by a consortium of the Naval Research Laboratory (USA), Max-Planck-Institut fur Aeronomie (Germany), Laboratoire d'Astronomie Spatiale (France), and the University of Birmingham (UK). SOHO is a project of international cooperation between ESA and NASA. In France, this work was supported by CNES.

\end{document}